\documentclass[twocolumn,showpacs,preprintnumbers,amsmath,amssymb]{revtex4}

\usepackage{graphicx}
\usepackage{dcolumn}
\usepackage{bm}

\begin{document}


\title{Formation of Dark Matter Haloes in a Homogeneous Dark Energy Universe}

\author{L. Marassi}

\email{luciomarassi@ect.ufrn.br}

\affiliation{Escola de Ci\^{e}ncias e Tecnologia\\ Universidade
Federal do Rio Grande do Norte, CEP 59072-970\\ Natal, Rio Grande
do Norte, Brazil}

\date{\today}

\begin{abstract}

Several independent cosmological tests have shown evidences that
the energy density of the Universe is dominated by a dark energy
component, which cause the present accelerated expansion. The
large scale structure formation can be used to probe dark energy
models, and the mass function of dark matter haloes is one of the
best statistical tools to perform this study. We present here a
statistical analysis of mass functions of galaxies under a
homogeneous dark energy model, proposed in the work of Percival
(2005), using an observational flux-limited X-ray cluster survey,
and CMB data from WMAP. We compare, in our analysis, the standard
Press-Schechter (PS) approach (where a Gaussian distribution is
used to describe the primordial density fluctuation field of the
mass function), and the PL (Power Law) mass function (where we
apply a nonextensive q-statistical distribution to the primordial
density field). We conclude that the PS mass function cannot
explain at the same time the X-ray and the CMB data (even at 99\%
confidence level), and the PS best fit dark energy equation of
state parameter is $\omega=-0.58$, which is distant from the
cosmological constant case. The PL mass function provides better
fits to the HIFLUGCS X-ray galaxy data and the CMB data; we also
note that the $\omega$ parameter is very sensible to modifications
in the PL free parameter, $q$, suggesting that the PL mass
function could be a powerful tool to constrain dark energy models.

\end{abstract}

\pacs{98.80.Es; 95.35.+d; 98.62.Sb}
\keywords{cosmology; structure formation; mass function.}
\maketitle

\section{Introduction}

In the last few years, several experiments and observations have
shown strong evidences that the expansion of our Universe is
accelerating, under the influence of a mysterious dark energy
component\cite{Observ,Sper07}. The dark energy presents an
equation of state parameter $\omega=p/\rho$ and represents about
74\% of the Universe. When $\omega=-1$ we have the special case of
the cosmological constant - or vacuum energy - as responsible for
the Universe acceleration (it is the standard $LCDM$ cosmological
model).

We also know that around $\sim$ 26\% of the Universe are composed
by matter (baryonic + dark matter); analyzing the growth of these
matter density fluctuation, we can describe the formation of the
large structures, such as clusters and super-clusters of galaxies.

The mass function of galaxies is a powerful tool to study the
large scale structure formation in the Universe. The standard
analytical approach was developed in 1974 by Press and
Schechter\cite{PS74} (hereafter PS), and it is still used today,
for its success and simplicity. The PS approach uses a Gaussian
distribution to describe the primordial matter density fluctuation
field.

Inspired by Tsallis $q$-nonextensive statistics\cite{Tsal88} and
kinetic theory\cite{RS00,ZER}, we have proposed a new mass
function\cite{L04,L07,L08}, replacing the Gaussian by a
non-Gaussian distribution - a Power Law (PL) distribution. The PL
mass function has a physically motivated free $q$ parameter
(related with the long range gravitational correlations between
particles), which provides malleability to fit observational data.
Also, if $q=1$ we recover the original Gaussian distribution. The
PL mass function is an extension of the PS one, and presents the
same analytical simplicity.

We have compared the PS mass function with the PL one, in the
special case of the $LCDM$ cosmological model\cite{L08}. Using the
X-ray flux-limited sample of galaxy clusters from Reiprich and
Boehringer\cite{reiprich02} (HIFLUGCS) in a $\chi^{2}$ statistical
analysis, and applying independent cosmological tests (BAO and
Shift Parameter) to better constrain the results, we concluded
that the PS approach presents incompatibilities with the
independent CMB data. On the other hand, using the PL mass
function, we have better cosmological parameters, and we note an
overlap with the CMB data for a large range of values of its $q$
free parameter. These results, although encouraging, are limited
to the $LCDM$ model ($\omega=-1$).

The mass function could be used to probe the dark energy behavior
in the Universe, and works using homogeneous and non-homogeneous
dark energy in the large scale structure formation are beginning
to flourish in the literature. In this work we use a model of
homogeneous dark energy from Percival\cite{Perci1}. By allowing
other values for the $\omega$ dark energy parameter, we can
compare the applicability of the PS and the PL mass functions to
probe dark energy models.

\section{The Mass Function Equations}
\label{Sec1}

\hspace{0.5cm} The distribution of bound objects with masses
between $M$ and $M+dM$, using the Gaussian distribution to
describe the primordial density fluctuation field
$\delta\equiv{\delta\rho}/{\rho}$ (The PS approach)
reads\cite{PS74}:
\begin{equation}
\frac{dF_{(M)}}{dM}=+\frac{1}{\sqrt{2\pi }}\frac{\delta
_{c}}{\sigma _{(M)}^{2}}\left( \frac{\partial \sigma
_{(M)}}{\partial M}\right) \exp \left( -\frac{\delta
_{c}^{2}}{2\sigma _{(M)}^{2}}\right)\label{fm3}
\end{equation}
where $\sigma _{(M)}^{2}\equiv \left\langle \delta
_{M}^{2}\right\rangle $ is the mean squared fluctuation, and
$\delta _{c}$ is the critical $\delta$ for collapse.

\hspace{0.5cm}Now, if instead of Gaussian we use the PL
distribution for the initial fluctuations, we derive the follow
expression\cite{L04,L07,L08}
\begin{eqnarray}
\frac{dF_{(M)_{PL}}}{dM} &=&+\frac{B_{q}}{\sqrt{2\pi }}\frac{\delta _{c}}{%
\sigma _{(M)}^{2}}\left( \frac{\partial \sigma _{(M)}}{\partial
M}\right)
\nonumber \\
&&\cdot \left[ 1-\left( 1-q\right) \cdot \left( \frac{\delta _{c}}{\sqrt{2}%
\sigma _{(M)}}\right) ^{2}\right] ^{\frac{1}{\left( 1-q\right) }}
\label{Fm-PL}
\end{eqnarray}
where the factor $B_{q}$ is a one-dimensional normalization
constant. In the limit $q\rightarrow 1$ the above PL expression
reduces to the standard Gaussian approach.

\begin{figure}
\includegraphics[width=2.2truein, height=2.35truein,
angle=-90]{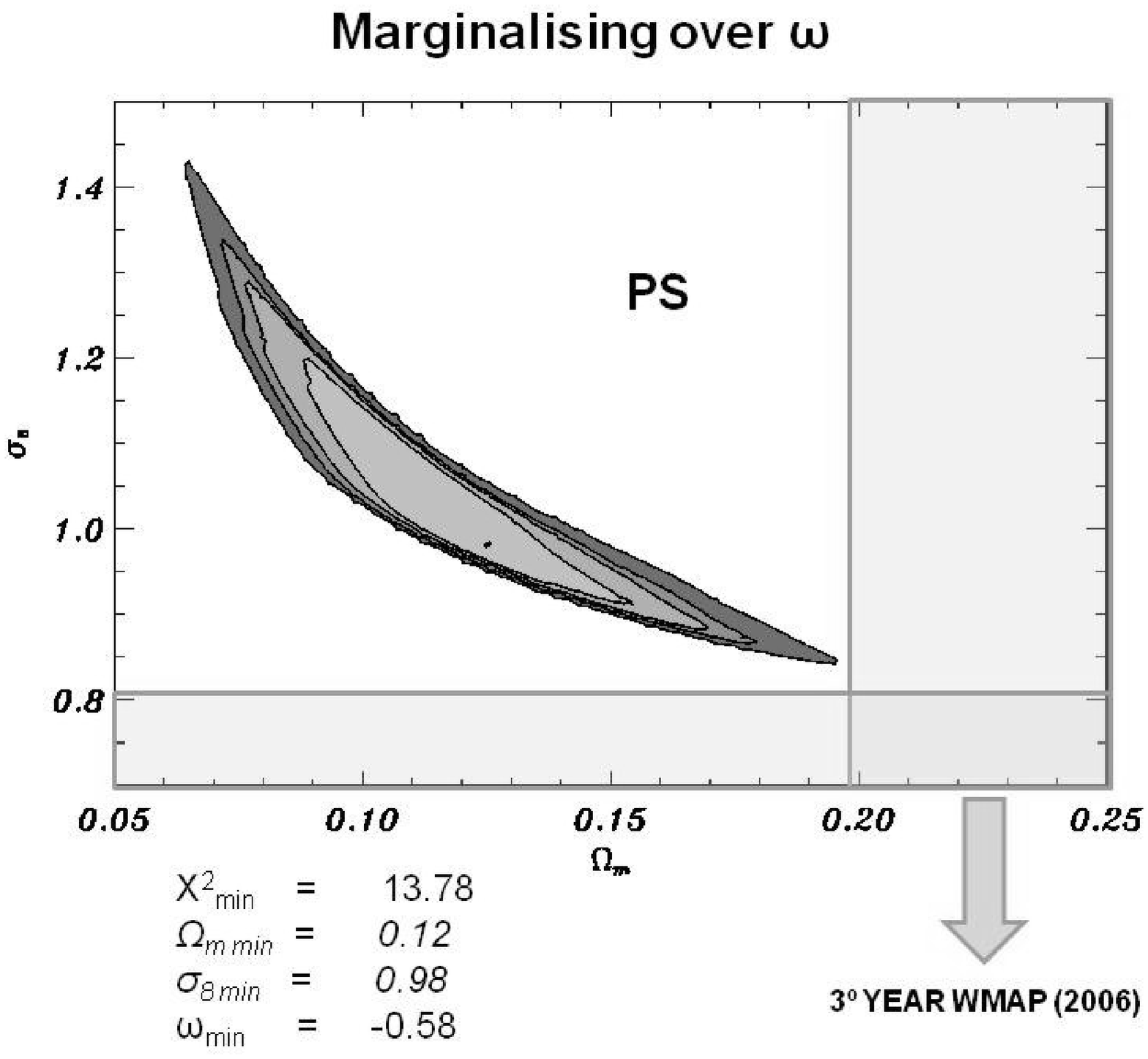}\,\,\,\,\,\,\,\,\,
\includegraphics[width=2.2truein, height=2.35truein, angle=-90]{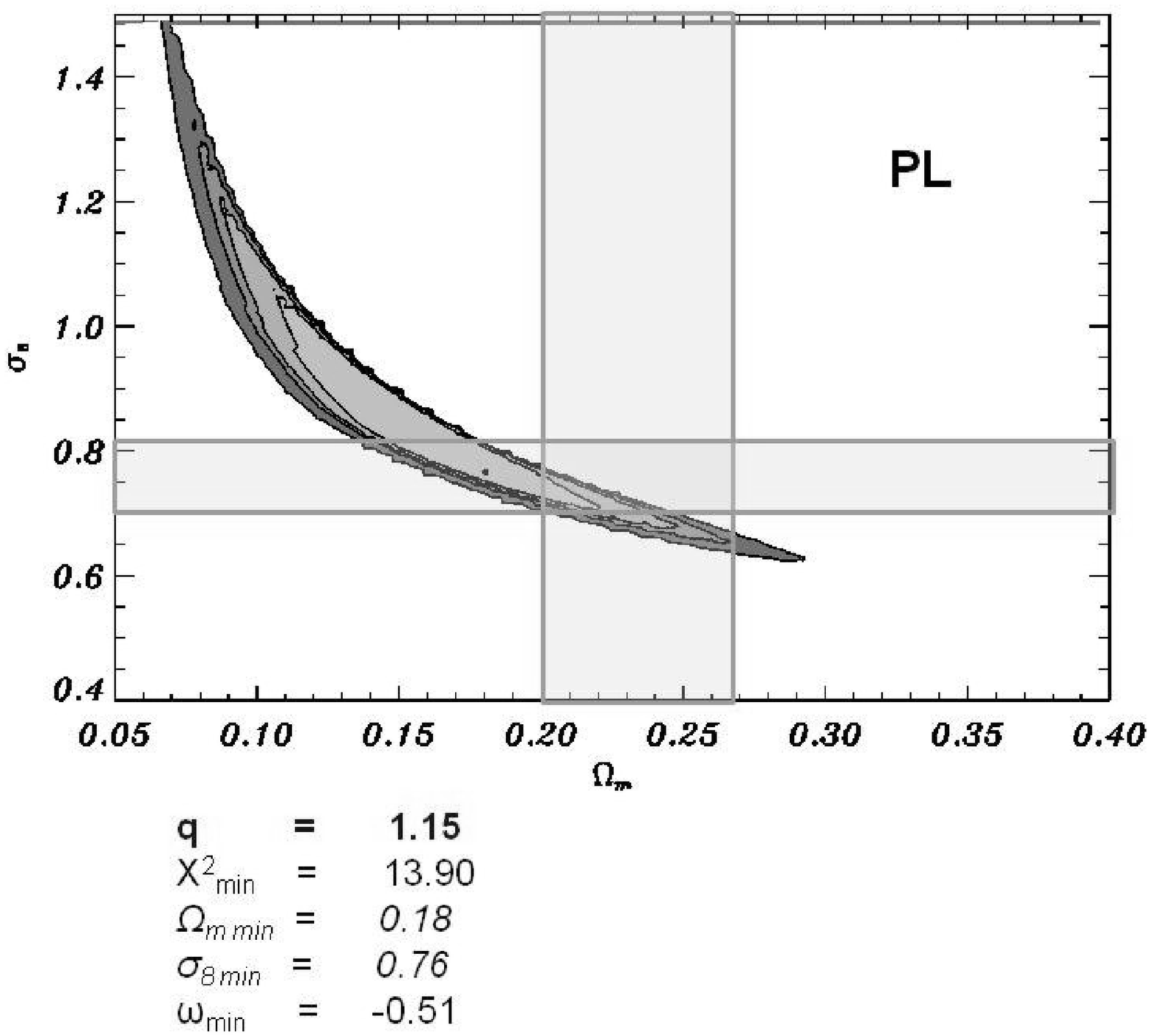}
\caption{Statistical contours fitting the mass function to the
HIFLUGCS X-ray data. We plot the $\Omega_{m}-\sigma_{8}$ plane,
marginalizing over the dark energy $\omega$ parameter. In both
panels, the solid vertical and horizontal lines show the minimum
and maximum WMAP limits for $\Omega_{m}$ and $\sigma_{8}$. Note
that the contours using the Gaussian distribution (the left panel)
does not intercept the WMAP values while the other one (in the
right panel), using the PL distribution with $q=1.15$, overlap the
WMAP independent data, for almost the same dark energy $\omega$
parameter.} \label{Fig1}
\end{figure}

\begin{figure}
\includegraphics[width=2.2truein, height=2.35truein,
angle=-90]{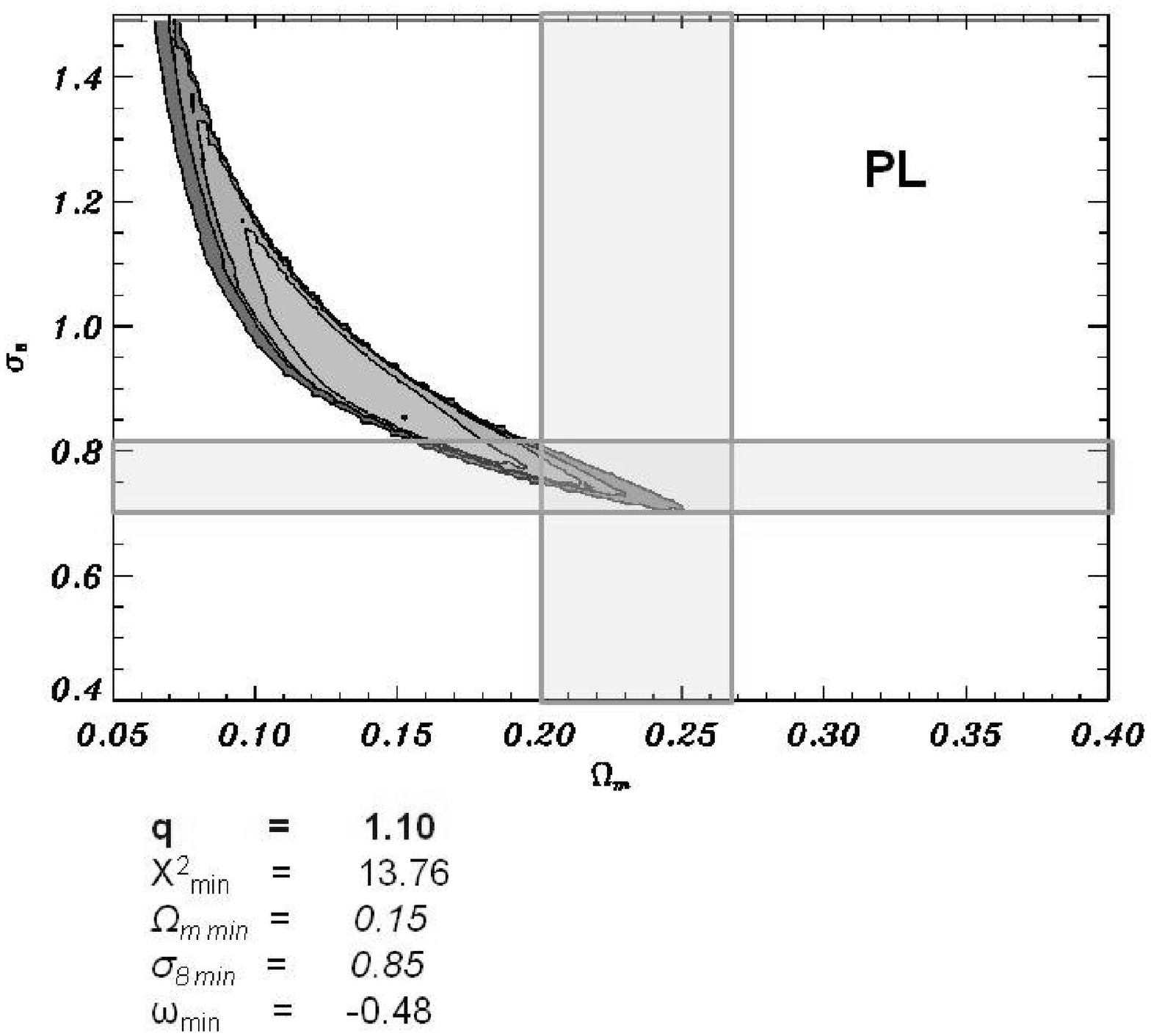}\,\,\,\,\,\,\,\,\,
\includegraphics[width=2.2truein, height=2.35truein, angle=-90]{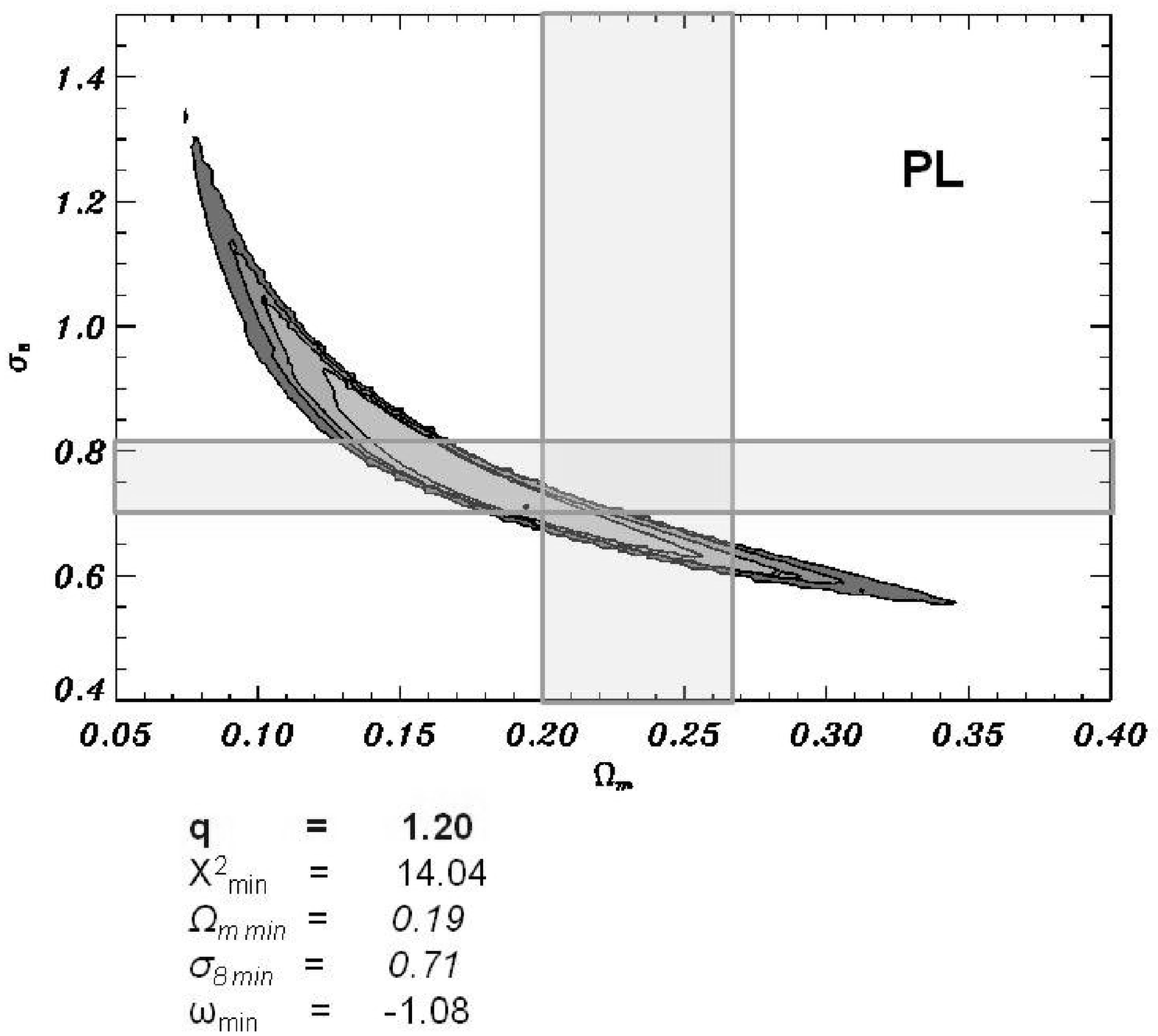}
\caption{Statistical contours on the $\Omega_{m}$-$\sigma_{8}$
plane, using the HIFLUGCS data with the PL mass function, under
the dark energy model of Percival (2005). Marginalizing over the
dark energy $\omega$, we see that the PL mass function overlap the
independent CMB data (solid lines), for a wide range of its free
parameter ($1.10<q<1.20$). Note that while $q$ grows from 1.10 to
1.20, the parameter $\omega$ presents huge modifications (from
-0.48 to -1.08) - the $\omega$ parameter is highly sensible to
changes in the PL mass function.} \label{Fig2}
\end{figure}

\section{Results and Future Perspectives}

We performed a $\chi^{2}$ statistical procedure using the X-ray
HIFLUGCS data sample\cite{reiprich02}, and so we determined the
statistical confidence contours (from $1\sigma$ to $3\sigma$
errors) for the pair of parameters, $\sigma_{8}$ and $\Omega_{m}$.
We applied the homogeneous dark energy model of
Percival\cite{Perci1}, and we marginalized over all possible
values of the dark energy equation of state parameter $\omega$.

In Figure \ref{Fig1} we show the  contours in the
$\Omega_{m}$-$\sigma_{8}$ plane obtained by using the PS and the
PL approaches. The left panel shows the PS results: the best-fit
to the HIFLUGCS X-ray galaxy cluster data are $\Omega_{m}=0.12$,
$\sigma_{8}=0.98$ and $\omega=-0.58$. The best fit values of
$\Omega_{m}$ and $\sigma_{8}$ are, respectively, very low and very
high when compared to the nowadays independent CMB
results\cite{Sper07}. the solid vertical and horizontal lines show
the minimum and maximum CMB limits (from WMAP 3 years data) for
the $\Omega_{m}$ and $\sigma_{8}$ parameters. The contours using
the Gaussian distribution (the left panel) does not intercept the
CMB values. In the other hand, by using the PL mass function with
a certain $q$ free parameter (the right panel of Figure
\ref{Fig1}), we  could overlap the WMAP independent data, for
almost the same dark energy $\omega$ parameter (in this example,
we use $q=1.15$). In Figure \ref{Fig1} we see that, for almost the
same particular model of dark energy ($\omega\sim-0.5$), we obtain
better cosmological parameters using the PL mass function, while
the standard PS mass function cannot explain the independent WMAP
data even at $99\%$ confidence level.

In Fig. \ref{Fig2}, we plot two statistical contours in the
$\Omega_{m}$-$\sigma_{8}$ plane, using the HIFLUGCS data with the
PL mass function, under the dark energy model of
Percival\cite{Perci1}. Again we marginalize over all possible
values of $\omega$. By fixing $q=1.10$ in the PL mass function, we
obtain as best-fit a dark energy $\omega=-0.48$, $\Omega_{m}=0.15$
and $\sigma_{8}=0.85$ (and the plot overlap the independent WMAP
data with this configuration). When we fix $q=1.20$ the plot
continues to overlap the WMAP data, with almost the same best-fit
for $\Omega_{m}$ and $\sigma_{8}$, but it presents a huge
modification in the best-fit dark energy $\omega$ parameter
($\omega=-1.08$ here, close to the cosmological constant case).
So, in the range $1.10<q<1.20$, the PL mass function provides
better fits to the HIFLUGCS X-ray galaxy data and the independent
data from CMB. Also, the sensibility over the $\omega$ parameter,
as we change the free $q$ parameter, indicates that the PL mass
function could be a powerful tool to constrain dark energy models.

It must to be noted, however, that we are using the standard Press
\& Schechter approach, in this present work. This approach smooths
the initial density fluctuations, and uses the spherical model to
calculate the epoch of critical overdensity for collapse of the
density perturbations ($\delta_c$ - critical linear density
contrast at collapse time). Such simple model eventually fail in
detail, given structure formation complexities like, for example,
the asymmetrical gravitational collapse; meanwhile, recent mass
function improvements from numerical simulations have now
quantified these problems - e.g. the Sheth \& Tormen ellipsoidal
collapse (1999 - hereafter ST)\cite{ST}, and the universal
function of Jenkins et al. (2001 - hereafter
Jenkins)\cite{Jenkins}.

In this work, we use the Percival critical overdensity for
collapse ($\delta_c$), with growth factor derived from the special
case of flat dark energy models with constant $\omega$ parameter
(equations 19 to 21 from reference \cite{Perci1}). This critical
overdensity is only weakly dependent on cosmological parameters,
and the cumulative mass function, using the numerical fit of ST,
presents very little difference for different cosmologies
(different $\omega$ dark energy parameters) at low redshifts ($z$
close to zero); but as we go further back in time (towards higher
redshifts), the difference becomes appreciable (figure 6 from
\cite{Perci1}). We see that the evolution of the mass function is
strongly dependent on $\omega$, due to the evolution of the linear
growth factor (which affects $\delta_c$ directly).

It is worth to discuss, at this point, the results from the work
of Matthew et al. (2009 - hereafter Matthew)\cite{Mat}. Matthew
studied the role of the linear density contrast at collapse time,
$\delta_c$, in Early Dark Energy (EDE) models - where dark energy
has a non-negligible importance since the beginning of the
structure formation process. Using a pure numerical approach to
compute $\delta_c$, Matthew found that, at redshift $z=0$, the EDE
mass functions are not greatly altered compared to the LCDM
cosmological model, but the difference increases with redshift -
the same conclusion as Francis et al. (2008, hereafter
Francis)\cite{Francis}. Matthew also demonstrates that the EDE
model presents a basic agreement between the ST and the Jenkins
mass functions, even if we use a cosmology dependent $\delta_c$,
derived from the spherical collapse model. However, when we
compare the Jenkins mass function (which is 'blind' to the growth
history of the universe) to the ST mass function with a fixed
$\delta_c$ ($\delta_c=1.689$), or deriving ST $\delta_c$ from the
spherical model (where the history of the universe - the growth
history - counts), we can see small differences between these mass
functions in the high mass end ($M\geq10^{13}M_\odot/h$); despite
these differences, the most up-to-date numerical simulations have
not sufficient accuracy at this high mass range to discriminate
between these two approaches.

So, we observe a general agreement between the conclusions of
Percival, Matthew and Francis, concerning dark energy models in
the structure formation. All dark energy mass functions, ST or
Jenkins, have almost the same behavior as the LCDM model at $z=0$,
and at high redshifts we can perceive differences between the
cosmological models, due to the evolution of the linear growth
function. In the work of Percival, the focus is the difference
based on the dark energy $\omega$ parameter, and in the work of
Matthew and Francis, the focus is the difference between the ST
(with fixed or evolving $\delta_c$) and the Jenkins mass function.

In this present work we use, as said before, the standard PS
approach, with a Gaussian and a non-Gaussian (PL) initial
distribution function. We use in this work a low-redshift X-ray
galaxy cluster survey, but even at small redshifts we show a
strong mass function dependence on the dark energy $\omega$
parameter, using the PL approach, and the $\chi^{2}$ statistical
procedure show significant contour differences between the dark
energy mass function and the LCDM cosmological model. This show
clearly the power of the PL mass function $q$ free parameter to
fit observational data; and if this malleability is observed to
galaxy clusters at $z\sim0$, we should expect even more exciting
results using high redshifts catalogs, to probe the
characteristics of the dark energy, in a close future. Also, we
intend to study these dark energy models using ST and Jenkins mass
functions as well, studying the $\delta_c$ influence in the
process, to compare our results with the previous conclusions of
Percival, Matthew and Francis.

\section{Conclusions}
\label{Sec3}

In this work we perform a statistical analysis, using the X-ray
galaxy cluster data named HIFLUGCS\cite{reiprich02}, and the
independent Cosmic Microwave Background radiation (CMB) data from
WMAP\cite{Sper07}, to study the influence of galaxy mass functions
under a homogeneous dark energy background, from the work of
Percival\cite{Perci1}. Performing a $\chi^{2}$ analysis and
marginalizing over the $\omega=p/\rho$ dark energy equation of
state parameter, we compare the standard Gaussian Press-Schechter
mass function\cite{PS74} to the PL mass function (based on the
non-extensive thermodynamics and kinetic theory), which presents a
physically motivated free parameter, $q$\cite{L04,L07,L08}.

We conclude that the PS mass function cannot explain at the same
time the X-ray and the CMB data, even at $3\sigma$ confidence
level, which imposes a problem to the standard Gaussian scenario.
Also, the PS approach presents a best fit dark energy equation of
state parameter of $\omega=-0.58$, which is clearly distant from
independent estimates of $\omega\sim-1$ (a cosmological constant).

Observing the contours of Figure \ref{Fig2}, in the range
$1.10<q<1.20$, the PL mass function provides better fits to the
HIFLUGCS and the CMB independent data, with almost the same
best-fit for $\Omega_{m}$ and $\sigma_{8}$, but with a huge
modification in the best-fit dark energy $\omega$ parameter, as we
change $q$. This strongly suggests that the PL mass function could
be a powerful tool to constrain dark energy models. Future
analysis, based on the next generation of cluster surveys in
medium and high redshifts, will certainly provide better
constraints to the dark energy model that accelerates de Universe.

\section*{Acknowledgements}
The author are grateful to ECT-UFRN and FAPESP, for the financial
support provided, and to Lima J.A.S. for helpful discussions.

\end{document}